\documentclass[runningheads]{llncs}

\usepackage[T1]{fontenc}
\usepackage{graphicx} % If possible, figure files should be included in EPS format.

\usepackage{sidecap} % allow side caption for narrow figures
\usepackage{subcaption}

\usepackage{amsmath}

\usepackage{lipsum}

\usepackage{booktabs}

\usepackage{array}
\newcommand{\PreserveBackslash}[1]{\let\temp=\\#1\let\\=\temp}
\newcolumntype{C}[1]{>{\PreserveBackslash\centering}p{#1}}
\newcolumntype{R}[1]{>{\PreserveBackslash\raggedleft}p{#1}}
\newcolumntype{L}[1]{>{\PreserveBackslash\raggedright}p{#1}}

\begin{document}

\title{Efficient simulation of non-uniform cellular automata with a convolutional neural network}
\titlerunning{Efficiently simulation of $\nu$CAs with a CNN}

\author{Michiel Rollier\inst{1}\orcidID{0000-0001-8467-734X} \and
Aisling J.~Daly\inst{1}\orcidID{0000-0002-3390-2495} \and Odemir M.~Bruno\inst{1}\orcidID{0000-0002-2945-1556} \and
Jan M.~Baetens\orcidID{0000-0003-4084-9992}}

\authorrunning{M.~Rollier et al.}

\institute{Univ.~Ghent, Dept.~Data Anal.~\& Math.~Modelling, BionamiX, Coupure Links 653, B-9000 Ghent, Belgium\\\email{michiel.rollier@ugent.be}
\and
Univ.~São Paulo, São Carlos Inst.~Phys., POB 369, BR-13560970 São Carlos, SP, Brazil}

\maketitle

\begin{abstract} % 150--250 words.
Cellular automata (CAs) and convolutional neural networks (CNNs) are closely related due to the local nature of information processing. The connection between these topics is beneficial to both related fields, for conceptual as well as practical reasons. Our contribution solidifies this connection in the case of non-uniform CAs ($\nu$CAs), simulating a global update in the architecture of the Python package \texttt{TensorFlow}. Additionally, we demonstrate how the highly optimised out-of-the-box multiprocessing in \texttt{TensorFlow} offers interesting computational benefits, especially when simulating large numbers of $\nu$CAs with many cells.

\keywords{Cellular automata \and Non-uniform \and Convolutional neural networks \and \texttt{CellPyLib} \and \texttt{TensorFlow}}
\end{abstract}

\section{Introduction}

\subsection{Elementary and non-uniform cellular automata}

Arguably the simplest non-trivial and maximally discrete dynamical system is an elementary cellular automaton (ECA). In this model, a finite or countably infinite number of cells are aligned in one dimension. A cell can be in only two possible states, all cells update their state in discrete time steps based on their own and their direct neighbours's state. Additionally, all cells update their states simultaneously, they do so deterministically, and they all follow the same local update rule (see e.g.~\cite{bhattacharjee2020survey}). Relaxing any of these conditions results in a CA that belongs to a family of discrete models that typically exhibit a richer behaviour, a more complex mathematical description, and well-defined `taxonomic' ties to other families. Our forthcoming comprehensive review on this taxonomy \cite{rollier2024comprehensive}) provides an overview.

\begin{figure}
    \centering
    \includegraphics[width=.618\linewidth]{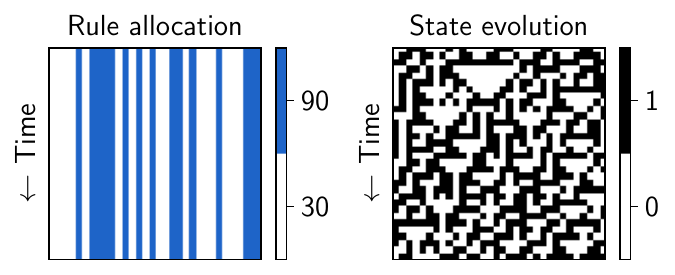}
    \caption{A non-uniform CA ($\nu$CA) is determined by the rules that govern the local update, the allocation of those rules (blue and white), and the initial configuration of the states (black and white). In this example, two rules $30$ and $90$ are allocated. The allocation (left) is such that the $\nu$CA is uniform in time but not in space. In combination with a particular initial configuration, this results in the state evolution shown on the right.}
    \label{fig:example-nuCAs}
\end{figure}

In particular, allowing certain cells to follow different local update rules results in the family of CAs collectively identified as non-uniform CAs ($\nu$CAs). Our review paper \cite{rollier2024comprehensive} covers non-uniformity in the most general sense, where the `rule allocation' varies in space and time. However, in the literature \cite{cattaneo2009nuca} $\nu$CA rule allocation is typically only spatially non-uniform. For this reason, together with the fact that our proposed implementation is more cumbersome in the general interpretation of a $\nu$CA, we will only consider spatially non-uniform CAs in this contribution. Additionally, as we will focus on simulating $\nu$CAs, we will only be concerned with finite grids. Fig.~\ref{fig:example-nuCAs} contains an example of a $\nu$CA with $N=32$ cells and $N_R = 2$ elementary rules.

Clearly, allowing non-uniformity implies that the space of possible CA dynamics increases in size considerably. An ECA with $N$-cells and periodic boundary conditions already has $2^N$ possible initial configurations, which evolves into different dynamics for each of the $256$ elementary rules. A $\nu$CA consisting of $N$ cells that each evolve according to one of $N_R$ rules has $2^N \times N_R^N$ such possible initial conditions. This large diversity obstructs mathematical generalisation except in particular cases that are quite remote from applications \cite{dennunzio2012nonuniform}. An empirical approach to phenomenological classification is therefore imperative, but such a computational task requires an efficient means of simulation.

\subsection{CA classification and simulation by means of CNNs}
\label{subsec:class_sim_cNNs}

The CA classification problem \cite{silverman2019convolutional,comelli2021comparing} is a challenge at the centre of CA research (see e.g.~\cite{vispoel2022progress}). Considering the fact that we can interpret the spacetime diagrams of CAs as images, computer vision techniques can be mobilised for their classification, including those researched in the domain of deep learning. Within the spectrum of deep learning, convolutional neural networks (CNNs) are wildly popular, largely due to their undeniable success in image processing and computer vision \cite{dhillon2020convolutional}. We refer to excellent monographs to gain a good understanding of the topic (e.g. \cite{goodfellow2016deep}), while a good visual introduction is offered by the deep learning series by 3Blue1Brown on YouTube \cite{3b1b2017deeplearning}.

A lot of diverse data is required in order to effectively train CNNs to identify classes. Fortunately, the local nature of the convolution operation enables not only the identification of CAs, but also their emulation. After all, nodes in a neural network may be identified with CA cells, and a convolutional operation can be interpreted as an update from a local neighbourhood. In fact, as Gilpin \cite{gilpin2019cellular} shows, the global update mechanism on any kind of CA can be accommodated by the architecture of a CNN. This can be achieved both by a clever choice of weights and biases, or by training the network from a random initialisation.

\begin{SCfigure}[1]
    \includegraphics[width=.49\linewidth]{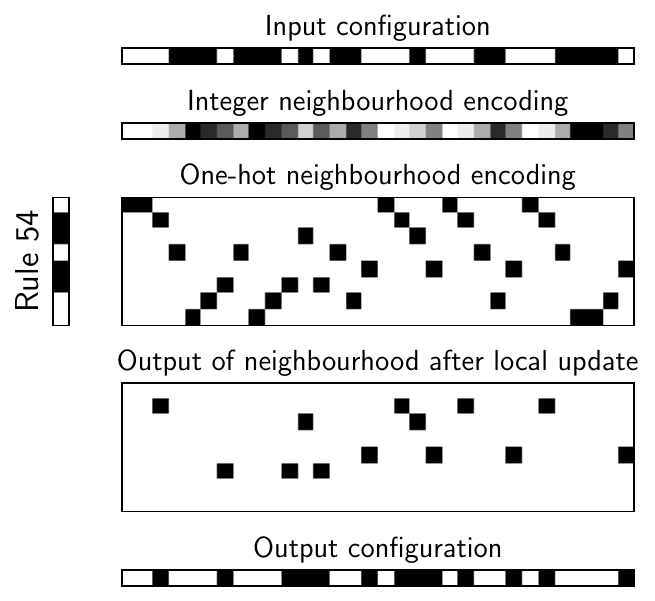}
    \captionsetup{skip=-12pt} % Adjust the vertical space here
    \caption{The subsequent (de)composition steps required for updating an ECA configuration, illustrated for $32$ randomly initialised cells, evolved over one time step by rule $54$. First, each binary size-$3$ neighbourhood is translated to an integer from $0$ to $7$ (shown in grey-scale). This integer is encoded as a size-$8$ one-hot vector (displayed in columns). Depending on the rule table of the local update rule (displayed on the left-hand side), this columns is kept or removed. As a final step, all columns are summed, resulting in the output configuration.}
    \label{fig:rule-decomposition}
\end{SCfigure}

In the CNN, transforming the input configuration to the neighbourhood encoding is performed by the first 1D convolutional layer, with a kernel of width $3$ and fixed weights $(4, 2, 1)$, zero bias, and periodic boundary conditions. The output of this convolution is transformed to a matrix with one-hot vectors as columns, and each of the $8$ rows of this matrix corresponds to a channel in the first CNN hidden layer. Next, another convolution layer with a kernel of width $1$ essentially sums all channels, where this time the weights are determined by the binary representation of the local update rule. The output is then, by design, the ECA configuration after one global update. With a mere $40$ parameters, this is an extremely simple CNN, whose computational complexity scales only with the number of cells $N$. The subsequent steps required to integrate a global update into a CNN framework are shown schematically in Fig.~\ref{fig:rule-decomposition}. This concrete example uses ECA rule $54$ and a random initial condition, but the required operations are independent of this choice

\begin{figure}
    \centering
    \includegraphics[width=.7\linewidth]{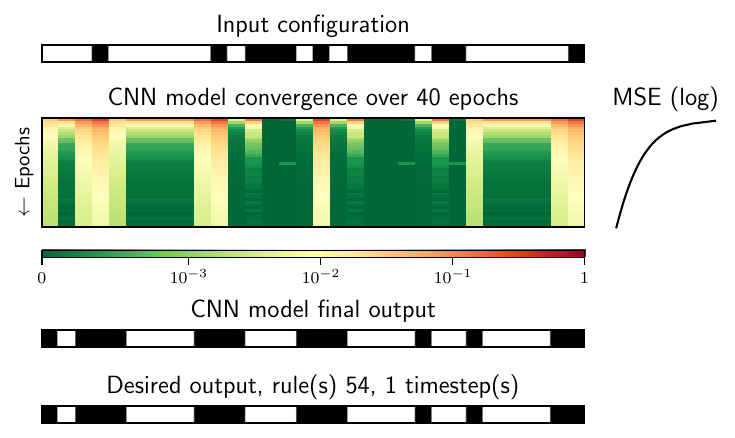}
    % \captionsetup{skip=-20pt} % Adjust the vertical space here
    \caption{A CNN can be trained to generate the desired next-timestep configuration of an ECA, illustrated here by showing the evolution of the inference on one of the training samples. Starting from a random choice of weights and biases, the mean square error (MSE) between the generated output and the desired output continues to shrink, while more and more epochs of data are mobilised in the training. This is shown in the qualitative MSE curve on the righthand side, and on a cell-per-cell basis in the log-scaled colour-coded heat map in the centre. For details on similar training procedures, consult \cite{gilpin2019cellular}.}
    \label{fig:plot_configs_ECA_32cells_rule54_40epochs_bs64_lr0p005}
\end{figure}

The parameters within the CNN (weights and biases) can be calculated, but for more general CAs they would typically be trained. In order for the CNN to be in practice (and consistently) trainable, starting from random weights and biases, some additional features are required. We will not focus on the training procedure here, but we may mention that the most important of such additional features would be activation functions \cite{goodfellow2016deep}. For illustrative purposes we include our convergence towards an optimum in parameter space for a CNN that emulates rule $54$ with near perfection in Fig.~\ref{fig:plot_configs_ECA_32cells_rule54_40epochs_bs64_lr0p005}. For details on preferred training procedures for CA emulators, we again refer to \cite{gilpin2019cellular}.

\subsection{Scope}

The goal of this article is to fill in a gap in the literature, by emulating $\nu$CAs by means of CNNs. We can benefit from the extremely streamlined software implementations designed for neural networks, optimised for parallel processing and general performance. That is to say: CNNs can present us with a practical tool for the fast and massive simulation of spacetime diagrams and analyses on these diagrams.

In the next section we will develop a CNN for $\nu$CAs, and we will see that this requires only a minimal addition to the architecture outlined above. We discuss some performance characteristics, and conclude with an outlook on the future of the marriage between CAs and CNNs.

\section{Methods}

In order to assess the performance of a CNN regarding the simulation of $\nu$CAs, we first discuss a popular well-established approach, and then introduce two varieties of CNN extensions.

\subsection{Existing approaches}

Some programming languages enable very convenient and computationally optimised ways of simulating and analysing CAs. Wolfram Mathematica is an obvious example, which was in fact partially created for this purpose \cite{gaylord1996modeling}. In Python, the most commonly used package is \texttt{CellPyLib} \cite{antunes2021cellpylib}.

\begin{SCfigure}[1]
    \centering
    \includegraphics[width=.55\linewidth]{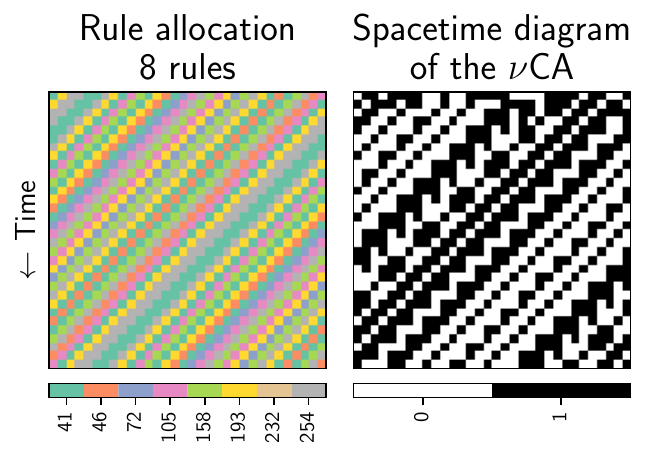}
    \captionsetup{skip=15pt} % Adjust the vertical space here
    \caption{A $\nu$CA is easily simulated using the Python package \texttt{CellPyLib}. One does so in two steps: first by defining the rule allocation (left, dependent on time and cell), and second by parsing this information to the \texttt{evolve} method, which generates the spacetime diagram (right).}
    \label{fig:cellpylib-spacetime_diagram-Nrules8}
\end{SCfigure}

It is straightforward to implement a $\nu$CA in \texttt{CellPyLib} by defining an array that instructs the \texttt{evolve} method on what rule to apply when to which cell. Adding more or fewer rules (i.e.~altering the non-uniformity) should not affect the performance. What does impact the performance, however, is the fact that the non-uniformity of the model no longer allows for caching the states in each step -- in \texttt{CellPyLib} this is encapsulated in the \texttt{memoize} option (sic). This means that one cannot make any `memory shortcuts' which typically speed up the CA simulation considerably. Fig.~\ref{fig:cellpylib-spacetime_diagram-Nrules8} displays an example for an $8$-rule $\nu$CA simulated in \texttt{CellPyLib}.

% Note that first transforming it to a uniform CA may also be relevant here (arguably much cheaper than the CNN approach).

\subsection{Two approaches for non-uniform CAs in \texttt{TensorFlow}}

Building on the CNN framework for elementary CAs, below we propose two additions which enable the CNN to emulate $\nu$CAs. Both additions involve a change in the CNN architecture between the one-hot neighbourhood encoding and the output layer. For technical details we refer to the code and annotations of the \texttt{NucaEmulator} class, available at mrollier/emulating-and-learning-CAs on GitHub, and entirely based on \texttt{TensorFlow} modules.

Note that the proposed CNNs emulate a single global update. Generating an entire spacetime diagram of $T$ time steps requires feeding the output of the CNN $T-1$ times back into the input layer, because in the proposed setup, the CNN can only emulate a single global update. Global updates strongly depend on the previous time step, so it is (in general) not possible to distribute the calculation of the CA dynamics into (for example) `all even time steps' and `all odd time steps'. This impossibility is related to the so-called computational irreducibility of CAs, and impedes temporal parallellisation of the computation of its dynamics. Note also that if the rule allocation is independent of time (as is conventionally the case for $\nu$CAs), the weights and biases of the CNN remain unchanged.

\subsubsection{A locally-connected hidden layer}

The first approach includes a locally connected layer, which is essentially a convolution where the kernel weights are allowed to differ for distinct nodes. Like in Fig.~\ref{fig:rule-decomposition}, all the columns are summed, weighted by the binary representation of the local update rule, but now the weights are not shared. The biases remain zero. The rest of the CNN is identical to that for ECAs. In practice one more intermediate step is added: our model first calculates the entire output configuration as if the CA would be uniform, for each of the $N_R$ rules. Next, the locally connected layer picks out the relevant cells based on which rule was actually allocated to it. While this is less computationally efficient, it does arguably increase interpretability of the model. More relevant in forthcoming research, however, is that this also facilitates flexibility in the training phase. After all, over-parameterisation is one of the key ingredients of deep learning.

Following the required subsequent calculations as explained in Section \ref{subsec:class_sim_cNNs}, this brings the total number of parameters in the model to
\begin{equation}
    (3+1) \times 8 + 8 \times N_R + N_R \times N,
\end{equation}
if we discard the bias parameters that have been set to zero. For the example depicted in Fig.~\ref{fig:cellpylib-spacetime_diagram-Nrules8}, this sums to $352$ parameters.

\subsubsection{A sparsely-populated dense layer}

A slightly different approach invokes the power of a fully-connected layer, known in the industry as a dense layer. Here, again CA outputs are calculated for each of the rules, but the cell selection now occurs by means of this dense layer. Note that, essentially, a locally connected layer is a dense layer for which all edges have been cut that connect nodes that represent different cells. Whilst this seems superfluous at first, there are two reasons to do so. First, \texttt{TensorFlow} is heavily optimised for calculating with large matrices, especially if these are sparse. Second, we again have the consideration of more model power and flexibility in future approaches that also involve training via backpropagation.

Technically, the $N_R$ channels containing size-$N$ outputs of the uniform case are first flattened, i.e.~deconstructed into a single vector of length $N_RN$. Next, all elements in this vector (the node values) are connected with the size-$N$ output layer by means of an $N_RN \times N$ weights matrix, where most weights are manually set to $0$ or $1$.

The total number of parameters in this model is therefore
\begin{equation}
    (3+1) \times 8 + 8 \times N_R + N_R \times N^2,
    \label{eq:cnn-loc-params}
\end{equation}
if again we do not count the vanishing bias parameters. For the example in Fig.~\ref{fig:cellpylib-spacetime_diagram-Nrules8}, this now sums to $8288$ parameters.

\subsection{Comparison of the three models in four scenarios}

% Before hinting at some of the more academic reasons for which bringing CA and CNNs closer together is exciting, let us first consider the mere computational benefit.
It is counter-intuitive that any increase in computing speed is expected at all, considering the clearly large increase in required floating point operations. To be fair, a really well-optimised and parallelised approach tailored to $\nu$CA simulation will undoubtedly outperform the proposed over-parameterised CNNs. The main allure, however, is in the combination of ease of use and out-of-the-box high performance of \texttt{TensorFlow} (or \texttt{PyTorch}, for that matter), as a result of the global scale of its continuous development.

We will briefly examine where the strengths and weaknesses of the CNN approaches lie, compared to the benchmark approach using \texttt{CellPyLib}. In particular we will consider four scenarios: the performance when adding more rules, more time steps, more cells, and more samples. Tab.~\ref{tab:scenarios} provides a summary of the parameter values (or ranges) that were found to be appropriate for best illustrating the trends and comparisons: these are the domains in which the overall trend in performance for all approaches is easily discernible. Every sample starts from a random initial condition but an identical rule allocation, such that the CNN needs to be initialised only once. Using the \texttt{time} Python package, we simply keep track of how many seconds each model requires for evolving the $\nu$CAs, taking the average over ten attempts.

\begin{table}
    \centering
    \caption{Four scenarios that enable the comparison of computational performance for $\nu$CA simulation by means of \texttt{CellPyLib}, and by means of the \texttt{TensorFlow} CNNs (locally connected and densely connected).}
    \begin{tabular}{L{.2\linewidth}L{.18\linewidth}L{.18\linewidth}L{.18\linewidth}L{.18\linewidth}}
        Scenario & Rules ($N_R$) & Time steps ($T$) & Cells ($N$) & Samples ($S$) \\ \toprule
        Alter $N_R$ & $[1, 256]$ & $32$ & $256$ & $32$ \\
        Alter $T$ & $4$ & $[10, 100]$ & $64$ & $32$ \\
        Alter $N$ & $4$ & $32$ & $[1, 256]$ & $32$ \\
        Alter $S$ & $4$ & $32$ & $32$ & $[1, 1024]$ \\ \bottomrule
    \end{tabular}
    \label{tab:scenarios}
\end{table}

This small-scale experiment was performed using an Intel Core i7-9850H CPU, 6 cores, at 2.6 GHz. We ran Python 3.11.8 and \texttt{TensorFlow} 2.14.0. Note, however, that the numerical value of the timing is secondary to the qualitative comparison.

\section{Results}

Here we present the computation times of the various scenarios listed in Table \ref{tab:scenarios}. We always show the \texttt{CellPyLib} data in blue, the data from the locally connected CNN in orange, and data from the fully connected CNN in green. In order to value the trends, we always show the mean value and the standard deviation from ten independent computations per unique combination of parameters.

\begin{figure}
    \centering
    \includegraphics[height=4.5cm]{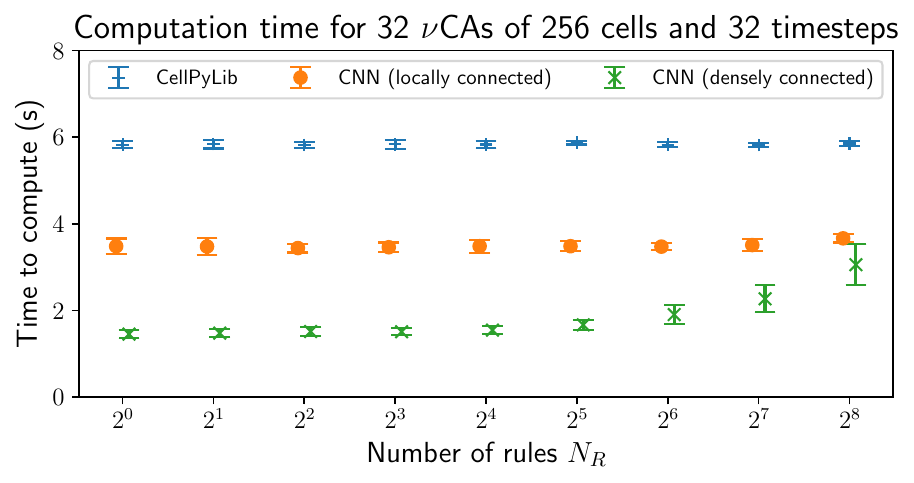}\\
    \includegraphics[height=4.5cm]{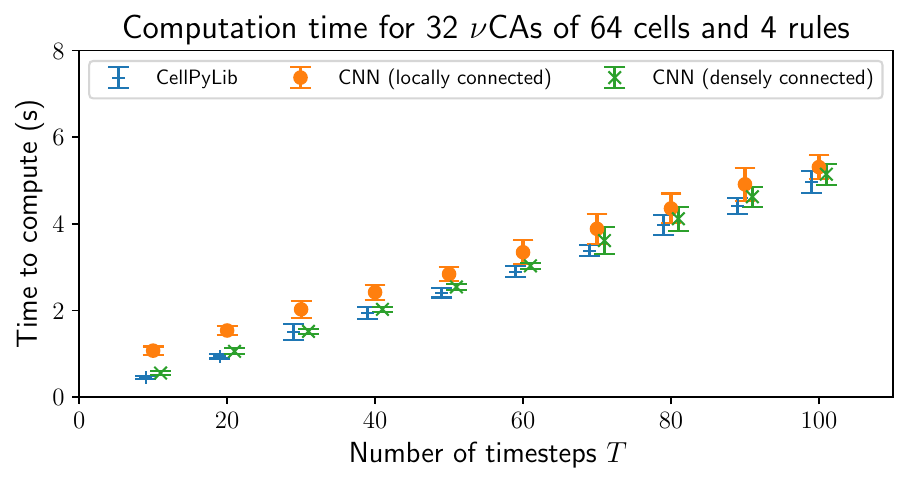}\\
    \includegraphics[height=4.5cm]{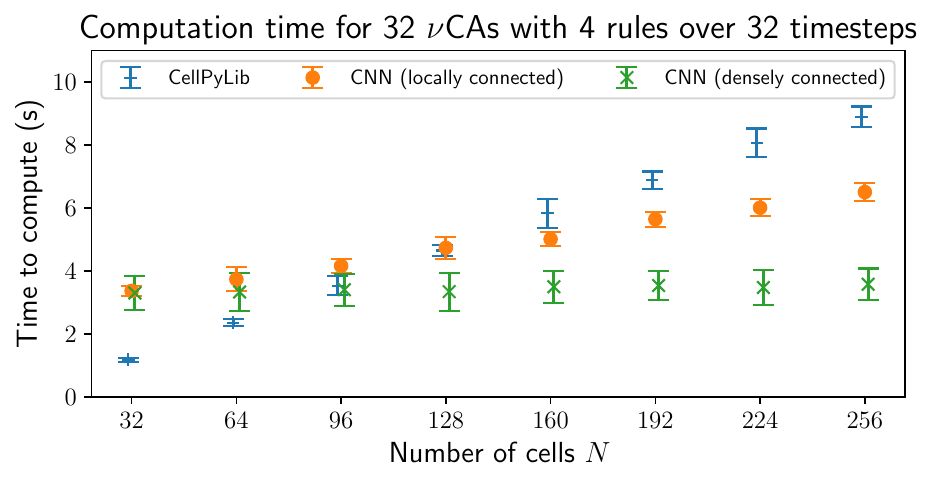}\\
    \includegraphics[height=4.5cm]{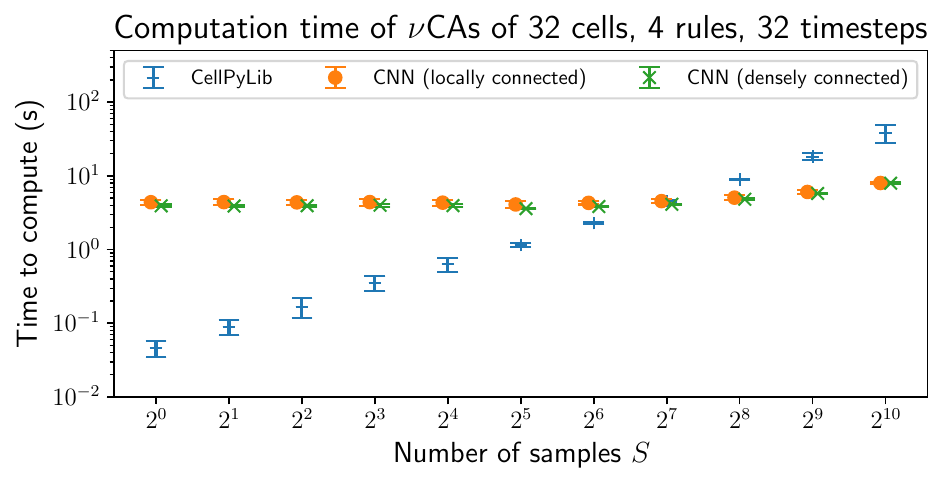}
    \caption{Results for all four scenario as listed in Tab.~\ref{tab:scenarios}, comparing three different simulation methods.}
    \label{fig:scenarios_results}
\end{figure}

Fig.~\ref{fig:scenarios_results} displays the computation times for the all four scenario. First, we show what happens when the number of rules (the `non-uniformity') is increased by factors of two. Because the number of rules $N_R$ goes up to $256$, we also chose $N=256$, allowing the possibility to allocate each rule at least once. We observe that the computation time is largely independent of $N_R$ for all models, except when a large number of rules is chosen in the densely connected CNN. Rather surprisingly, however, for small values of $N_R$ this densely connected CNN is significantly smaller than the other two.

For the second scenario, the number of time steps increases linearly between $10$ and $100$, and the computation time in all models appear to increase linearly as well. Except for small $T$ values, the computation time is similar for all three models.

The results from the third scenario are shown for eight linearly spaced values of $N$ between $32$ and $256$. As expected, \texttt{CellPyLib}'s computation time is proportional to the number of cells. The locally connected CNN also increases more or less linearly -- but with a higher `start-up cost'. The computation time of the densely connected CNN is largely independent of $N$.

For the fourth and final scenario, we consider $11$ logarithmically spaced values of $S$ between $1$ and $1024$. We again observe that using the larger models requires a certain initial cost, but once we want to simulate a large number of diagrams, they are clearly the least time consuming option.

% \subsubsection{A note on dynamic rule allocation}

% [this is where the current setup will supposedly fail most]

\section{Discussion, conclusion and prospects}

The highly optimised `out-of-the-box' multiprocessing of \texttt{TensorFlow} is clearly preferred over \texttt{CellPyLib} in scenarios where we want to generate many samples of $\nu$CAs with many cells. This of course is precisely the condition for obtaining statistically significant results in empirical studies of these discrete dynamical systems, especially when training models for automatic classification.

More surprising, however, is that the densely connected CNN almost always beats the locally connected CNN, despite the fact that mathematically speaking, the latter is a subgraph of the former. This is even the case when simulating more cells, despite, as Eq.~\eqref{eq:cnn-loc-params} shows, the quadratic growth of the number of parameters. This precisely demonstrates the point: \texttt{TensorFlow} is so cleverly optimised, that more complex models can outperform the simpler ones. This is arguably the reason why the \texttt{LocallyConnected1D} layer is discontinued in more recent versions of \texttt{TensorFlow}.

CNNs are a great tool for efficient simulation, which enables a more thorough exploration of the computational landscape of CAs. Similar approaches will enable the simulation of other types of CAs. As we show in forthcoming work, for example, graph CNNs are quite straightforward to mobilise in the simulation of network automata as well. While we do not claim that \texttt{TensorFlow} is the computationally optimal solution for CA simulation, it does present the CA researcher with an educational, ergonomic and flexible engine for efficient simulation.

CNNs are more than a tool for studying CAs, however. Arguably the most promising possibilities are created when, inversely, CAs serve the theoretical study and practical applications of CNNs. The training of CAs in the CNN framework may be used to better understand the information flow and learning process of CNNs \cite{gilpin2019cellular}. Another exciting avenue is the mobilisation of CAs for generative neural networks, as was elegantly illustrated in \cite{mordvintsev2020growing}. In any case, increased efforts in joining discrete dynamical modelling and deep learning, such as the one shared in this work, offer interesting benefits for both research domains.

\begin{credits}
% \subsubsection{\ackname} This study was funded by X (grant number Y).

\subsubsection{\discintname}
The authors have no competing interests to declare that are
relevant to the content of this article.
\end{credits}

\flushleft
\bibliographystyle{splncs04}
\bibliography{submission-acri2024}

\begin{thebibliography}{10}
\providecommand{\url}[1]{\texttt{#1}}
\providecommand{\urlprefix}{URL }
\providecommand{\doi}[1]{https://doi.org/#1}

\bibitem{3b1b2017deeplearning}
3Blue1Brown: Deep learning (Oct 2017), \url{https://www.youtube.com/watch?v=aircAruvnKk}

\bibitem{antunes2021cellpylib}
Antunes, L.M.: Cellpylib: A python library for working with cellular automata. Journal of Open Source Software  \textbf{6}(67), ~3608 (2021). \doi{10.21105/joss.03608}, \url{https://doi.org/10.21105/joss.03608}

\bibitem{bhattacharjee2020survey}
Bhattacharjee, K., Naskar, N., Roy, S., Das, S.: A survey of cellular automata: types, dynamics, non-uniformity and applications. Natural Computing  \textbf{19}(2),  433--461 (6 2020). \doi{10.1007/s11047-018-9696-8}

\bibitem{cattaneo2009nuca}
Cattaneo, G., Dennunzio, A., Formenti, E., Provillard, J.: Non-{Uniform} {Cellular} {Automata}. vol.~5457, pp. 302--+ (2009). \doi{10.1007/978-3-642-00982-2_26}

\bibitem{comelli2021comparing}
Comelli, T., Pinel, F., Bouvry, P.: Comparing {Elementary} {Cellular} {Automata} {Classifications} with a {Convolutional} {Neural} {Network}. In: {ICAART} (2). pp. 467--474 (2021)

\bibitem{dennunzio2012nonuniform}
Dennunzio, A., Formenti, E., Provillard, J.: {Non-uniform cellular automata: Classes, dynamics, and decidability}. Information and Computation  \textbf{215},  32--46 (2012). \doi{https://doi.org/10.1016/j.ic.2012.02.008}

\bibitem{dhillon2020convolutional}
Dhillon, A., Verma, G.: Convolutional neural network: a review of models, methodologies and applications to object detection. Progress in Artificial Intelligence  \textbf{9}(2),  85--112 (Jun 2020). \doi{10.1007/s13748-019-00203-0}

\bibitem{gaylord1996modeling}
Gaylord, R.J., Nishidate, K.: Modeling nature: Cellular automata simulations with Mathematica{\textregistered}. Springer (1996)

\bibitem{gilpin2019cellular}
Gilpin, W.: Cellular automata as convolutional neural networks. Physical Review E  \textbf{100}(3),  032402 (Sep 2019). \doi{10.1103/PhysRevE.100.032402}

\bibitem{goodfellow2016deep}
Goodfellow, I., Bengio, Y., Courville, A.: Deep {Learning}. MIT Press (2016)

\bibitem{mordvintsev2020growing}
Mordvintsev, A., Randazzo, E., Niklasson, E., Levin, M.: Growing {Neural} {Cellular} {Automata}. Distill  (2020). \doi{10.23915/distill.00023}

\bibitem{rollier2024comprehensive}
Rollier, M., Zielinski, K.M.C., Daly, A.J., Bruno, O.M., Baetens, J.M.: A comprehensive taxonomy of cellular automata (2024)

\bibitem{silverman2019convolutional}
Silverman, E.: Convolutional {Neural} {Networks} for {Cellular} {Automata} {Classification} (Jan 2019). \doi{10.1162/isal_a_00175}

\bibitem{vispoel2022progress}
Vispoel, M., Daly, A., Baetens, J.: {Progress, gaps and obstacles in the classification of cellular automata}. Physica D: Nonlinear Phenomena  \textbf{432}, ~30 (2022), \url{http://hdl.handle.net/1854/LU-8740550}

\end{thebibliography}

\end{document}